\documentclass[twocolumn,aps,prl,groupedaddress,amssymb]{revtex4}
\usepackage{natbib}
\usepackage{graphicx}
\usepackage{units}
\usepackage{braket}
\usepackage{amsmath}
\usepackage{dsfont}

\begin{document}
\hyphenation{Ryd-berg}

\title{Hybridization of Rydberg electron orbitals by molecule formation }

\author{A. Gaj}
\email[Electronic address: ]{a.gaj@physik.uni-stuttgart.de}
\author{A. T. Krupp}
\author{P. Ilzh\"{o}fer}
\author{R. L\"{o}w}
\author{S. Hofferberth}
\affiliation{5. Physikalisches Institut and Center for Integrated Quantum Science 
and Technology, Universit\"{a}t Stuttgart, Pfaffenwaldring 57, 70569 Stuttgart, 
Germany}
\author{T. Pfau}
\email[Electronic address: ]{t.pfau@physik.uni-stuttgart.de}
\affiliation{5. Physikalisches Institut and Center for Integrated Quantum Science 
and Technology, Universit\"{a}t Stuttgart, Pfaffenwaldring 57, 70569 Stuttgart, 
Germany}
\date{11 March 2015}
\begin{abstract}
The formation of ultralong-range Rydberg molecules is a result of the attractive 
interaction between Rydberg electron and polarizable ground state atom in an 
ultracold gas. In the nondegenerate case the backaction of the polarizable atom
on the electronic orbital is minimal. Here we demonstrate, how controlled 
degeneracy of the respective electronic orbitals maximizes this backaction and 
leads to stronger binding energies and lower symmetry of the bound dimers. 
Consequently, the Rydberg orbitals hybridize due to the molecular bond.

\end{abstract}
\maketitle
\begin{figure*}[htb]
\includegraphics[width=160mm]{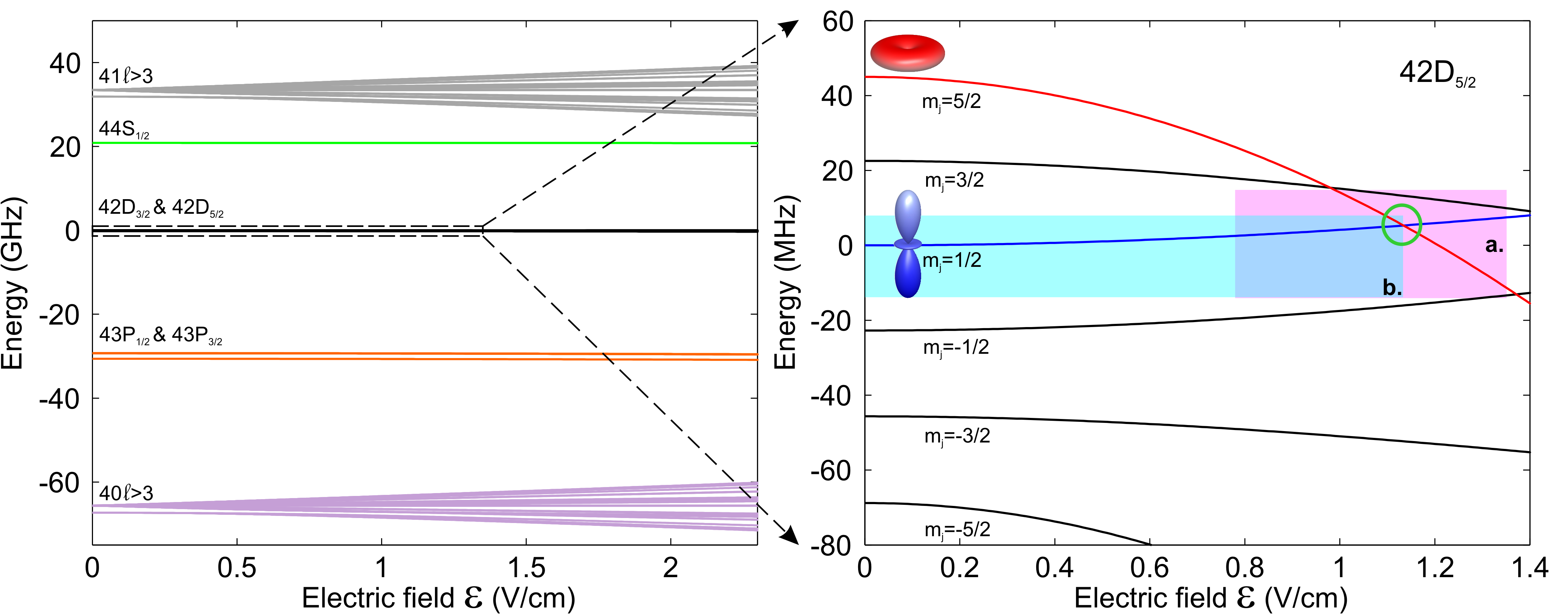}
\caption{ Calculated Stark map in the vicinity of the 42$D$ state for $^{87}$Rb. 
The electric field  is aligned parallel to the magnetic field $B$=\unit[13.55]{G}. 
All energies are plotted relative to the energy of the 42$D_{5/2}, m_j=1/2$ state 
at $\mathcal{E} $=0. The states are labeled with quantum numbers of the states 
they adiabatically unite with at $\mathcal{E} $=0. In the magnification the sixfold 
splitting of the 42D$_{5/2}$ is shown. The green circle marks the investigated 
crossing of the $m_j$=1/2 (blue) and $m_j$=5/2 (red) energy levels. The part of 
the spherical harmonics that is relevant for the molecule formation at $\mathcal{E} $=0 
is depicted close to the corresponding levels. The shaded areas a. (magenta) and b. 
(light blue) correspond to the measurement ranges of Fig. \ref{Fig2}.}
\label{Fig1}
\end{figure*}
\begin{figure}[b!]
\includegraphics[width=85mm]{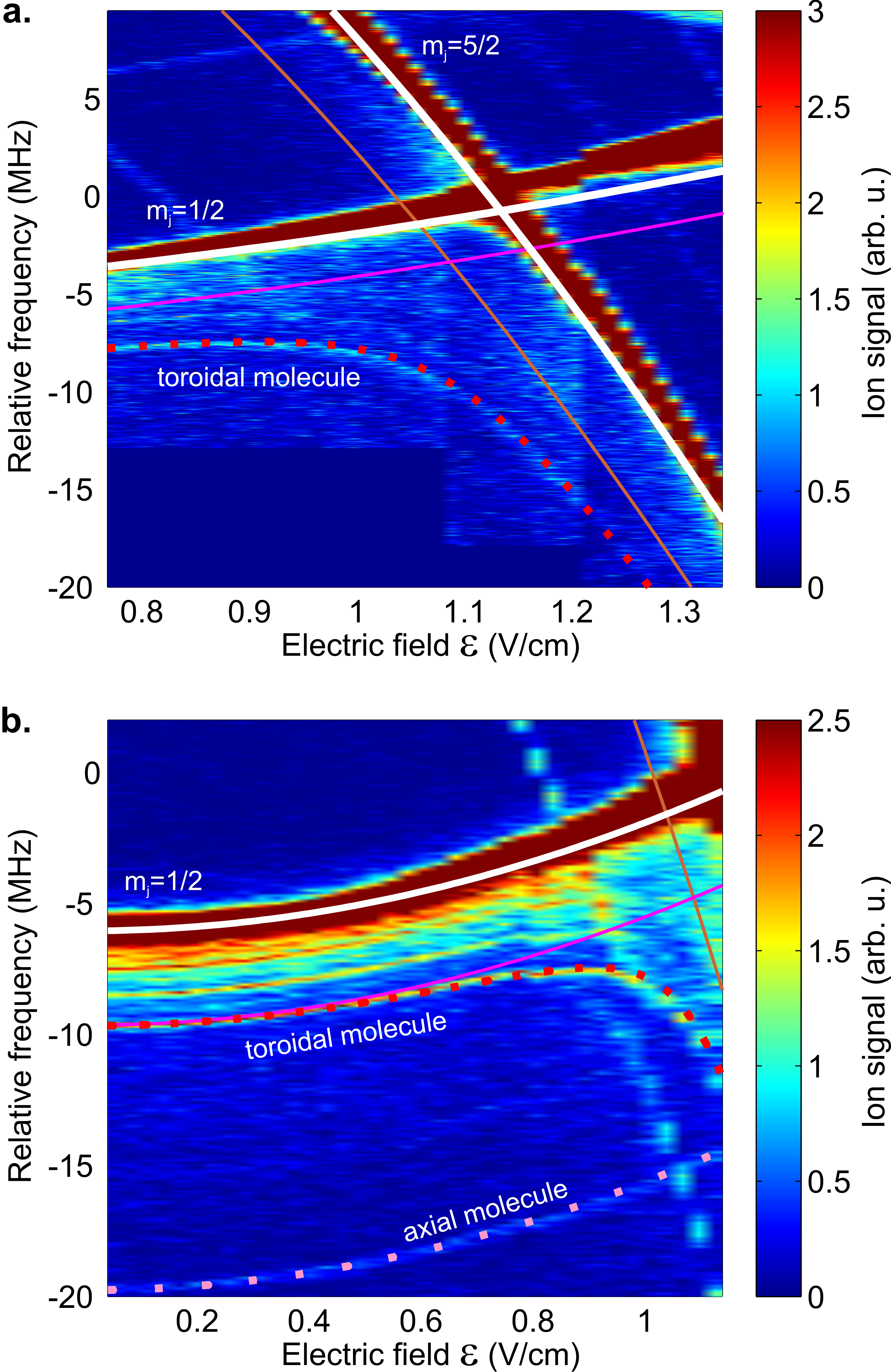}
\caption{Stark map in the vicinity of (a.) the degeneracy of the two atomic lines 
m$_j$=5/2 and m$_j$=1/2 and (b.) the m$_j$=1/2 state followed from the zero 
electric field up to the crossing. Computed atomic states positions, shown before 
in Fig. \ref{Fig1} are depicted with continuous white lines. All theoretical 
molecular lines are plotted with respect to these atomic energy levels. The 
hypothetical positions of the ground state toroidal molecular states, with no 
hybridization of the electron orbital, are indicated with brown and magenta lines, 
for m$_j$=5/2 and m$_j$=1/2, respectively. The binding energy of the toroidal 
molecule belonging to the m$_j$=1/2 state changes in the vicinity of the atomic 
crossing. Calculated positions of the ground state toroidal and axial molecules 
are indicated, for the sake of visibility, with red and pink dotted lines, 
receptively. In addition excited rovibrational toroidal states are visible in b. 
However, it is not possible to accurately trace them for higher electric fields. 
We identify the weak diagonal lines visible at higher electric field as laser 
sidebands with no physical meaning. Note that the atomic lines are saturated due 
to the chosen color scale. The discontinuity in the m$_j$=1/2 atomic line above 
\unit[1.2]{V/cm} is an experimental artifact.}
\label{Fig2}
\end{figure}
\begin{figure}[t!b]
\includegraphics[width=86mm]{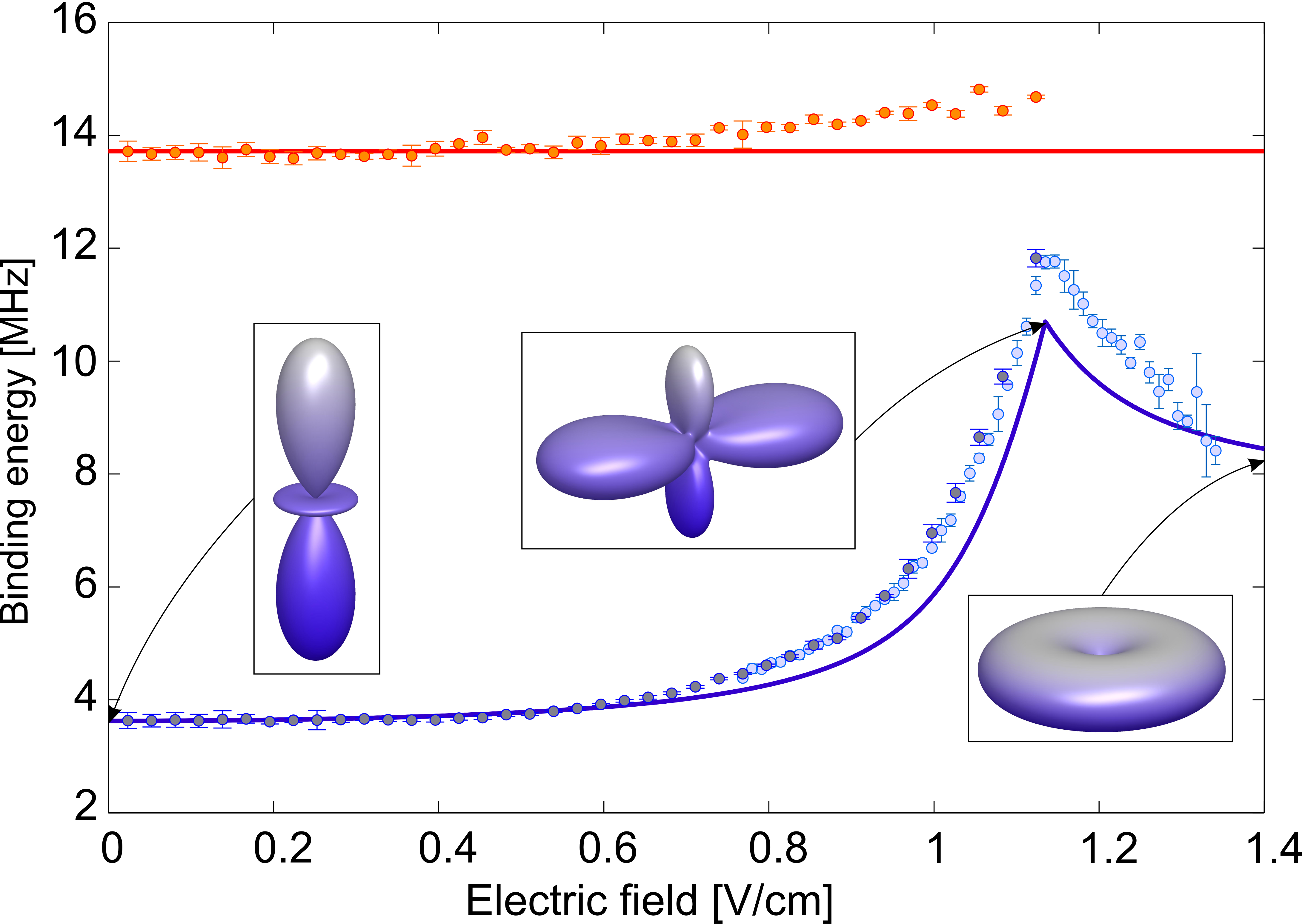}
\caption{Measured binding energies of the axial (orange points) and toroidal 
ground state molecules (light blue and violet points) versus the electric field. 
The binding energies are extracted from two data sets shown in Fig. \ref{Fig2} 
(a. and b.) and hence indicated with two colors for the toroidal molecule data. 
A ground state atom inside the Rydberg electron in the toroidal plane mixes two 
orbitals. As a result the binding energy of the m$_j$=1/2 toroidal molecule 
increases, reaches its maximum directly at the crossing and then asymptotically 
decreases to the E$_B$ value of m$_j$=5/2 toroidal molecule. The relevant part 
for molecule formation of the spherical harmonic of the hybridized Rydberg orbital 
at the crossing is shown in the inset together with the two asymptotic shapes. 
The error bars are determined as two standard deviations of the fit. }
\label{Fig3}
\end{figure}

The geometrical structure of molecules determines their physical and chemical 
properties. The shape of an individual molecular bond can be explained by the 
concept of hybridization. The mixing of nearly degenerate atomic orbitals leads 
to a new hybrid orbital, which allows to analyze structures of many basic 
molecules like carbon dioxide, ammonia or water. One of the first molecules 
described this way is methane \cite{Pauling1931}. In CH$_4$ the carbon 
electrons, initially in $s$ or $p$ configuration, are rearranged into four hybrid 
orbitals $sp^3$ while creating the molecular bond. The new hybrid orbital contains 
25\%  of $s$ character and 75\% of $p$ character.
\\ \indent
Here, we present the hybridization of a Rydberg electron orbital induced by the 
formation of an ultralong-range Rydberg molecule. Rydberg molecules consist of a 
single Rydberg atom bound to one or more ground state atoms. The bond in this type 
of molecules originates from the elastic scattering between the slow Rydberg 
electron and a ground state atom. Diatomic Rydberg molecules have been observed 
for S-states \cite{Bendkowsky2009,Tallant2012}, D-states \cite{Krupp2014,Anderson2014} 
and P-states \cite{Bellos2013,Heiner2014} in gases of Rb or Cs. Polyatomic Rydberg
molecules have been recently photoassociated from a gas of Rb \cite{Gaj2014}. 
The presence of the neutral perturber inside the Rydberg electron orbital can 
cause admixing of a nearly degenerate hydrogenic manifold and energetically close 
by $l$-states at the location of the ground state atom \cite{Greene2000}. As a 
result, these so-called trilobite molecules, can possess a giant permanent electric 
dipole moment. For Cs, which has an almost integer quantum defect for the S-state, 
this admixture can amount to 90\% and lead to a permanent dipole moment of 
kilo-Debye, which was shown in \cite{Booth2014} for principal quantum numbers 
$n$=37, 39, 40. In Rb, the hydrogenic manifold is energetically much further away. 
Therefore the hydrogenic admixture was less then 1\% and the observed permanent 
dipole moments were on the order of 1 Debye for $n$=35, 43 \cite{Li2011}. In this 
paper, unlike in the experiments with trilobite states, we do not mix many states 
with different principal and angular quantum numbers. We demonstrate the perturber 
induced formation of a new hybrid orbital, which is a linear combination of only 
two m$_j$-states with known orbital shapes. For this reason, it makes it a 
textbook-like example of tunable hybridization in the field of Rydberg molecules. 
The ratio of the contributing states can be precisely controlled by an applied 
electric field. This allows for tuning the contribution of one m$_j$-state to the 
hybrid orbital from 0 to 100\% over the crossing of the two m$_j$-states. 
\\ \indent
In the absence of an external field, the magnetic substates of the Rydberg energy 
levels are degenerate. This degeneracy can be lifted by applying a magnetic 
field B. The initially degenerate states splits then into $2j+1$ different Zeeman 
levels labeled by the projection of the total angular momentum on the quantization 
axis $m_j$. Individual $m_j$-states can be brought back to degeneracy by 
applying an additional electric~field~$\mathcal{E}$ aligned parallel to the 
B-field. In Fig. \ref{Fig1} calculated Stark maps in the vicinity of the 42$D$ 
state in an external magnetic field of B=\unit[13.55]{G} are shown.
The eigenenergies are obtained by diagonalizing the full Hamiltonian consisting 
of an unperturbed atomic Hamiltonian and two terms accounting for the interaction 
with the magnetic and the electric field. The splitting between states of different 
quantum numbers $l$, with $l\leq 3$  is caused by their quantum defects 
\cite{Li2003,Mack2011}. These states exhibit a quadratic Stark effect, as can be 
seen for the 42D$_{5/2}$ state, which splits into six $m_j$ levels, shown in the 
magnification of Fig. \ref{Fig1}. States of the manifold with $l>3$ show a linear 
Stark effect, which results in a fan-like Stark structure.  The magnitude of the 
corresponding shifts depends on the absolute value of $m_j$. Since the electric 
and magnetic dipole operators do not couple states of $\Delta m_j \neq 0$ for 
$\mathcal{E} \parallel$B, the  $m_j$=5/2 and $m_j$=1/2 states become degenerate 
at  $\mathcal{E} $=\unit[1.135]{V/cm} exhibiting real crossing. The absence of 
coupling between the atomic states provides a clean two level system, suitable for the 
investigation of hybridization of the electronic orbitals due to perturber induced 
coupling.
\\ \indent
To describe the interaction of a Rydberg atom with a ground state atom in an
external homogeneous electric and magnetic field, we write down the Hamiltonian:
\begin{equation}
H_{tot}=H_0+H_B+H_\mathcal{E} +\frac{\textbf{P}^2}{M}+V_s(\textbf{r},\textbf{R}),
\label{eq:1}
\end{equation}
where $\frac{\textbf{P}^2}{M}$ represents the kinetic energy of the ground state perturber in a center of mass frame, where $M$ is the reduced mass,
$H_0$ is a field-free electron Hamiltonian and $H_B$ and $H_\mathcal{E} $ account 
for the interaction with the electric and magnetic field, respectively. In this 
approach the total three body system is described by two relative positions 
\textbf{R} of the ground state atom and \textbf{r} of the electron, taken with 
respect to the Rydberg ionic core as well as the respective momenta \textbf{P} 
and \textbf{p}. Additionally, particles are treated as point-like. The pseudopotential 
$V_s(\textbf{r},\textbf{R})$ describing the interaction between 
the low-energy electron and the perturbing atom takes, in the $s$-wave regime, the form 
\cite{Fermi1934}
\begin{equation}
V_s(\textbf{r},\textbf{R})=\frac{2 \pi \hbar^2 a_s}{m_e} \delta(\textbf{r}-\textbf{R}),
\label{eq:2}
\end{equation}
where $a_s$ is the $s$-wave scattering length and $m_e$ is the electron mass. 
For $^{87}$Rb the triplet scattering length is negative and as a consequence 
the potential (\ref{eq:2}) is attractive. Taking into account the local electron 
density, the resulting potential is proportional to 
$\left| \Psi (\textbf{R}) \right|^2$ \cite{Greene2000}. Its modification due to 
the $p$-wave shape resonance can result in a binding by internal quantum reflection 
\cite{Bendkowsky2010} and in the creation of butterfly-type molecules 
\cite{Hamilton2002}. Momentum-dependent corrections to the scattering length can 
be calculated using a semiclassical approximation \cite{Omont1977}. We apply a 
Born-Oppenheimer approximation and express the total wavefunction as 
$\Psi(\textbf{r},\textbf{R})=\psi(\textbf{r},\textbf{R})\phi(\textbf{R})$,
where $\psi(\textbf{r},\textbf{R})$ is the electronic molecular wavefunction in 
the presence of the perturber and $\phi(\textbf{R})$ describes the molecular 
rovibrational state. This allows for computing the adiabatic potential energy 
surfaces (APES) for the fine structure states. Subsequently, one can calculate 
the molecular binding energies and the rovibrational molecular wavefunctions by 
solving the Schr\"{o}dinger equation using the previously calculated APES. This 
method was used before for theoretical studies of molecules in external magnetic 
and electric fields  \cite{Lesanovsky2006, Kurz2013, Kurz2014} and was experimentally 
verified in \cite{Krupp2014}. 
\\ \indent
Here, we are interested in the mixing of two orbitals $m_j$=5/2 and $m_j$=1/2 
due to the presence of the neutral atom inside the Rydberg electron orbit, which 
manifests itself in a change of the molecular binding energy E$_B$. We treat the 
molecular potential $V_s(\textbf{r},\textbf{R})$ as a perturbation in first order 
to the Hamiltonian $H'=H_0+H_B+H_\mathcal{E}$. 
For the calculation of the full Hamiltonian we use the basis of 
42D$_{5/2}$, $m_j$=5/2 and 42D$_{5/2}$, $m_j$=1/2 electron wavefunctions
\begin{equation}
\begin{split}
&\psi_{1/2}(\textbf{r})=R_{42D}(\textbf{r})\, Y_{1/2}(\Theta)=\sqrt\frac{3}{5}R_{42D}(\textbf{r})\, Y_2^0(\Theta) \\
&\psi_{5/2}(\textbf{r})=R_{42D}(\textbf{r})\, Y_{5/2}(\Theta,\phi)=R_{42D}(\textbf{r})\, Y_2^2(\Theta,\phi), 
\end{split}
\label{eq:3}
\end{equation}
where we include the respective Clebsch-Gordan coefficients and the spherical 
harmonics (shown in Fig. \ref{Fig2}) in accordance to triplet state of the electron 
involved. The radial wavefunctions are the same for both states. Since we take the 
unperturbed wavefunctions as a basis, there is no dependence on \textbf{R} left. 
Consequently, we obtain the full Hamiltonian
\begin{equation}
H =  \eta
 \begin{pmatrix}
 |Y_{1/2}|^2+\Delta_1/\eta	 &	 Y^*_{1/2}Y_{5/2} \\
 Y_{1/2}Y^*_{5/2} 	 & 	|Y_{5/2}|^2+\Delta_2/\eta
 \end{pmatrix},
\label{eq:4}
\end{equation}
where $\eta=\eta(\textbf{R})=\int{d\textbf{r} \, V_s(\textbf{r},\textbf{R}) \, |R_{42D}(\textbf{r})|^2}$
and $\Delta_1$ and $\Delta_2$ are the summed-up energy shifts arising from $H'$.
By diagonalizing the Hamiltonian (\ref{eq:4}) we obtain the new eigenenergies
of the system. The degeneracy between the basis states occurs for $\Delta_1=\Delta_2$.
In the experiment we can control the energy difference $\Delta_1=\Delta_2-\Delta_2$
and thus the states mixing by tuning the electric field.
\\ \indent
We excite Rydberg atoms in a two-photon process from a magnetically 
trapped spin-polarized ultracold cloud of $^{87}$Rb atoms in the $5S_{1/2}, F=2,m_F=2$ 
state \cite{LWN12} with an applied magnetic offset field of B=\unit[13.55]{G}. The 
typical temperature of the cloud is \unit[2]{$\mu$K} and the peak density is on the order 
of \unit[10$^{12}$]{cm$^{-3}$}. The 780nm laser drives the lower transition. The detuning 
from the intermediate 5P$_{3/2}$ state is \unit[500]{MHz}. The 480nm laser 
driving the upper transition is on constantly. After each \unit[50]{$\mu$s} long 
780nm laser pulse we ionize Rydberg atoms and molecules with an electric field and 
detect the ions on a microchannel plate detector. We perform 800 cycles of 
excitation, ionization and detection in one cloud while scanning the frequency of 
the red laser. This way we obtain one full spectrum of the Rydberg signal for a 
given value of the electric field. We adjust the polarization of the light driving 
the upper transition such that the intensities of the two atomic lines close to 
the crossing (Fig.~\ref{Fig1}) are comparable.  
\\
\indent
In Fig.~\ref{Fig2} we show the measured Stark maps in the vicinity of the 
m$_j$=5/2 and m$_j$=1/2 atomic states' crossing, marked in Fig.~\ref{Fig1} by the 
green circle. Due to stray electric fields present in the experiment, the electric 
and magnetic field are not perfectly parallel. The angle between $\vec{\mathcal{E}}$ 
and $\vec{B}$ is on the order of a few degrees, which in principle results in a very small coupling between the two atomic states. Nevertheless,  given the resolution 
of our experiment, the considered atomic states appear as a true crossing. The molecular 
states are shifted to the red with respect to the atomic line. This energy 
difference corresponds to the molecular binding energy E$_B$. Therefore, we can 
identify the molecular states by comparing their binding energies with the 
calculated values \cite{Krupp2014} from tracing them back to the zero electric 
field (Fig.~\ref{Fig2}b). The molecular state visible in the Fig.~\ref{Fig2}a 
originates from a ground state toroidal molecule bound in the equatorial plane 
($\Theta=\pi/2$) of the m$_j$=1/2 scattering potential (see inset of Fig.~\ref{Fig1}) 
with E$_B$=\unit[3.6]{MHz} at $\mathcal{E}$=0. The molecular ground state bound 
in the axial lobe ($\Theta=\pi,0$) of the same scattering potential occurs at 
E$_B$=\unit[13.7]{MHz}. This axial molecule follows in parallel the trace of the  
m$_j$=1/2 atomic line up to the degeneracy point, whereas the toroidal bends 
around the crossing  and asymptotically follows the m$_j$=5/2 atomic line. In the 
experiment we are able to address the individual rovibrational molecular states 
visible in the Fig.~\ref{Fig2}b. However, since the excited toroidal states are 
blurred around the atomic degeneracy, we focus our analysis on the molecular 
ground states.  
\\ \indent
Assuming no hybridization, the molecular states of the m$_j$=1/2 and m$_j$=5/2
would occur at a constant binding energy E$_B$($\mathcal{E}$=0) with respect to 
the corresponding atomic lines regardless of the applied electric field. 
In Fig.~\ref{Fig2} such unperturbed ground state toroidal molecule traces, plotted 
in reference to the calculated atomic energy levels, are shown for the m$_j$=1/2 
and m$_j$=5/2 states with continuous brown and magenta lines, respectively. These 
two states, described by the wavefunctions (\ref{eq:3}), form a basis for our two 
level model. 
Consequently to predict the behavior of the perturbed hybrid states, we calculate 
the eigenenergies E$_\pm$ of the Hamiltonian (\ref{eq:4}). For a given molecular 
state, the obtained analytical solution depends only on the energy difference 
$\Delta$ between the m$_j$=1/2 and m$_j$=5/2 states and a parameter $\eta$, which 
is determined from the fitting of E$_+$ to the experimental data at $\mathcal{E}$=0. 
In the experiment the energy difference $\Delta$ around the crossing depends to 
a good approximation linearly on the electric field. E$_-$, which would correspond 
to the toroidal molecule of the m$_j$=5/2 state was not observed in this experiment. 
By calculating E$_+$, i.\,e. the binding energy, as a function of $\Delta(\mathcal{E})$, 
we reproduce the behavior of the toroidal and axial molecular lines (Fig. \ref{Fig2})
with high accuracy. Note that the modeled molecular traces are plotted with respect 
to the computed atomic positions. The scattering potential couples the unperturbed 
toroidal molecular lines, although the respective atomic states cross without any 
level repulsion between them. This leads to orbital mixing and therefore a change 
in the Rydberg electron probability density, which we observe as an increased 
binding energy of the toroidal molecule. In case of the axial molecule, the off 
diagonal coupling terms in the Hamiltonian (\ref{eq:4}) vanish. This is due to 
the fact that the torus-like m$_j$=5/2 spherical harmonic is zero along the axial 
direction and thus does not modify the m$_j$=1/2 state in this direction. For 
this reason the binding energy stays constant.
\\ \indent
In Fig.\,\ref{Fig3} the measured and calculated (E$_+$) binding energies of the 
considered molecules versus the electric field are shown. Here the binding energy is determined as the difference between the measured atomic and molecular energy. The binding energy of 
the toroidal molecules increases from E$_B$($\mathcal{E}$=0) of the m$_j$=1/2 
state, reaches its maximum of E$_B$=\unit[11.7]{MHz} and asymptotically decreases 
to E$_B$($\mathcal{E}$=0) of the m$_j$=5/2 state. This transition can be attributed 
to an increase of the admixture of the m$_j$=5/2 state from 0 to 100\%. The 
maximum binding energy corresponds to an orbital consisting of 50\% m$_j$=1/2 
character and 50\% m$_j$=5/2 character. The resulting shape of the orbital is 
shown in the inset of Fig.\,\ref{Fig3}. The hybridization introduces a 
$\phi$-dependence to the Rydberg orbital, even though the asymptotic states were 
spherically symmetric along the $z$-axis. For the unperturbed axial molecule, the 
shape of the Rydberg orbital remains the same. 
\\ \indent
Directly at the crossing the experimental binding energy of the toroidal molecule 
and the calculated one deviate by 9\%. The predominant reason of the discrepancy 
around the crossing, for $\mathcal{E}>\unit[0.8]{V/cm}$, are difficulties in 
determining the atomic peak position caused by its non-Gaussian shape. Additionally, 
for $\mathcal{E}>\unit[1.2]{V/cm}$ fitting of the molecular state position becomes 
problematic. Moreover, for the calculation we only take the unperturbed states 
and not the real Stark states into account. Finally, more levels could be included 
into the analysis. In view of these facts, the agreement between the experimental 
data and our simple model is remarkable. A more complicated analysis 
would not lead to a significantly better agreement due to the purely experimental 
constrains.
\\ \indent
We have shown the hybridization of the Rydberg electron orbital due to the molecule
formation around the crossing of the atomic lines m$_j$=1/2 and m$_j$=5/2. The backaction of the bound perturber increases the binding energy and consequently changes the shape of the electron orbital.
Mixing more than two electron orbitals  of known shapes by bringing them to degeneracy
could result in even more complex asymmetric electron configurations. Furthermore, 
the influence of a different type of perturber, like an atom of another element 
or a heteronuclear molecule \cite{Rosario2014}, as well as the effect of a few 
perturbers on the electron orbital could be investigated. Rydberg electron orbital
hybridized due to thousands of the ground state perturbers could be observed
as an imprint on a Bose Einstein condensate \cite{Karpiuk2014}. Finally, engineering a state 
by placing the perturber precisely at a desired position might also be feasible 
in experiments with individual atoms in microtraps. 
																																				
\begin{acknowledgments}
We acknowledge support from the Deutsche Forschungsgemeinschaft (DFG) within the 
SFB/TRR21 and the project PF 381/13-1. Parts of this work was also funded by the
ERC under contract number 267100. A.G. acknowledges support from the E.U. Marie Curie 
program ITN-Coherence 265031 and S.H. from DFG through the project HO 4787/1-1. \\
A.G. and A.T.K. contributed equally to this work.
\end{acknowledgments}


\end{document}